\documentclass[a4paper,11pt]{article} 
\pdfoutput=1 
\usepackage{jcappub} 
\usepackage[T1]{fontenc} 

\newcommand{\be}{\begin{equation}}
\newcommand{\ee}{\end{equation}}
\newcommand{\simless}{\lower.5ex\hbox{$\; \buildrel < \over \sim\;$}}
\newcommand{\simgreat}{\lower.5ex\hbox{$\; \buildrel > \over \sim\;$}} 
\newcommand{\mion}{m_{\rm ion}}
\newcommand{\mbar}{\langle m \rangle} 

\newcommand{\thetacen}{\Theta_{\rm c}} 

\newcommand{\tcent}{T_{\rm c}}

\newcommand{\rgas}{{\cal R}} 
\newcommand{\mzero}{\mu_0} 
\newcommand{\conlum}{{\cal C}} 
\newcommand{\bcon}{f_{\rm g}}  
\newcommand{\mpro}{m_{\rm p}}
\newcommand{\mplan}{M_{\rm P}} 
\newcommand{\rplan}{R_{\rm P}} 
\newcommand{\tplan}{T_{\rm P}} 
\newcommand{\melect}{m_{\rm e}} 
\newcommand{\sigmasb}{\sigma_{\rm sb}} 
\newcommand{\effish}{{\cal E}} 
\newcommand{\starmass}{ M_0 }

\title{\boldmath Constraints on Alternate Universes: 
Stars and habitable planets with different fundamental constants} 

\author[a,b]{Fred C. Adams}

\affiliation[a]{Physics Department, University of Michigan, Ann Arbor, MI 48109} 
\affiliation[b]{Astronomy Department, University of Michigan, Ann Arbor, MI 48109} 

\emailAdd{fca@umich.edu} 

\abstract{This paper develops constraints on the values of the  
fundamental constants that allow universes to be habitable. We focus
on the fine structure constant $\alpha$ and the gravitational
structure constant $\alpha_G$, and find the region in the
$\alpha$-$\alpha_G$ plane that supports working stars and habitable
planets. This work is motivated, in part, by the possibility that
different versions of the laws of physics could be realized within
other universes. The following constraints are enforced: [A]
long-lived stable nuclear burning stars exist, [B] planetary surface
temperatures are hot enough to support chemical reactions, [C] stellar
lifetimes are long enough to allow biological evolution, [D] planets
are massive enough to maintain atmospheres, [E] planets are small
enough in mass to remain non-degenerate, [F] planets are massive
enough to support sufficiently complex biospheres, [G] planets are
smaller in mass than their host stars, and [H] stars are smaller in
mass than their host galaxies. This paper delineates the portion of
the $\alpha$-$\alpha_G$ plane that satisfies all of these constraints.
The results indicate that viable universes --- with working stars
and habitable planets --- can exist within a parameter space where the
structure constants $\alpha$ and $\alpha_G$ vary by several orders of
magnitude. These constraints also provide upper bounds on the structure
constants ($\alpha,\alpha_G$) and their ratio. We find the limit
$\alpha_G/\alpha\simless10^{-34}$, which shows that habitable
universes must have a large hierarchy between the strengths of the
gravitational force and the electromagnetic force. }

\begin{document}
\maketitle
\flushbottom

\section{Introduction} 
\label{sec:intro} 

A long standing problem is that the laws of physics are described by a
collection of fundamental constants, but we have no definitive
explanation for how the measured values of these constants are
determined \citep{dirac,gamow}. One partial explanation is provided by
the possible existence of other universes \citep{guth,vilenkin}, where
these separate regions of space-time could display variations in the
laws of physics, and hence variations in the fundamental constants. In
this scenario, the values of the constants in a given universe are
drawn from a set of underlying probability distributions. Our universe
represents one particular realization, i.e., one choice for the values
of the constants. Unfortunately, we do not have a theory for
predicting the form of the underlying probability distributions.  In
fact, there is not even a general concensus on what fundamental
parameters should be allowed to vary from universe to universe (for
example, compare the various suggestions presented in
\citep{reessix,hogan,tegold,aguirre,tegmark,burost,schell}).

An important issue is that the fundamental constants in our universe
have the proper values to allow life to develop. On the other hand,
different values could result in a lifeless universe, one with no
observers \citep{carr,bartip,reessix}. The necessity for us to live in
a universe with observers thus provides a partial explanation for why
the fundamental constants have their experimentally measured values.
In order to make this type of explanation more complete, however, we
need to know [i] what variations of the laws of physics are possible,
[ii] the probability distributions for the allowed variations, and
[iii] what subset of the possible universes allow for observers.  Many
authors have suggested that the universe is ``fine-tuned'' for the
development of life, i.e., that relatively small changes in the laws
of physics would preclude the development of observers
\citep{bartip,reessix,hogan}. However, the definition of what
constitutes fine-tuning is not the same for all authors and remains
unsettled (a detailed discussion is given in \citep{barnes}).

One basic step toward a resolution of the aforementioned issues is to
determine what values of the fundamental constants allow for the
existence of astrophysical structures, such as planets, stars, and
galaxies. The goal of this paper is to provide a partial answer.
Specifically, we consider constraints placed on a subset of the
fundamental constants by requiring that a universe support the
existence of both working stars and habitable planets. Parts of this
issue have been addressed previously using a variety of approaches
\citep{bartip,reessix,hogan,tegold,aguirre,tegmark,burost}. In
addition, previous work has considered the formation and structure of
galaxies in relation to habitable planets \citep{tegrees,coppess}.  As
a rule, however, these previous papers adopt a general approach, e.g.,
by considering a wide range of parameter space and relying on order of
magnitude arguments. In contrast, this paper uses solutions to the
equations of stellar structure (following the formalism of
\citep{adams}) to estimate the range of allowed stellar masses,
stellar lifetimes, and stellar surface temperatures across the range
of parameter space. This paper thus extends previous work by
presenting a more detailed treatment of stellar structure. Finally, we
note that the consideration of universes with different values of the
fundamental constants is related to the problem of time variations of
the constants in our universe \citep{uzan}.

It will be useful to work in terms of the structure constants defined 
through the relations 
\be
\alpha \equiv {e^2 \over \hbar c} \qquad {\rm and} \qquad 
\alpha_G \equiv {G \mpro^2 \over \hbar c} \, ,
\ee
where $\mpro$ is the proton mass.  In our universe, with standard
values of the fundamental constants, these parameters have the values
$\alpha \approx 1/137$ and $\alpha_G \approx 5.91 \times 10^{-39}$. 
In this paper, we consider a parameter space in which these structure
constants are allowed to vary by ten orders of magnitude in either
direction.

The weakness of gravity relative to the other forces (small values of
$\alpha_G/\alpha$) is an important aspect of the hierarchy problem in
particle physics, where it is more natural for the strengths of the
forces to be comparable. On the other hand, a small value of the ratio
$\alpha_G/\alpha$ is necessary for stars to exist \citep{adams}, and
for stars to have sufficiently long lifetimes and hot surface
temperatures (see Section \ref{sec:stellarcon}). Additional
constraints on the ratio $\alpha_G/\alpha$ arise from the ordering of
mass scales of planets, stars, and galaxies (as emphasized in earlier
work \cite{rees1972,reessix}, and constrained further in Section
\ref{sec:ordering}). This work thus shows that habitable universes
must display an enormous hierarchy of force strengths. 

For future reference, we also define the fundamental stellar mass 
scale $\starmass$ according to 
\be 
\starmass \equiv \alpha_G^{-3/2} \mpro = \left( {\hbar c \over G} 
\right)^{3/2} \mpro^{-2} \, \approx \, 
3.7 \times 10^{33} {\rm g} \approx 1.85 M_\odot \,,
\label{mscale} 
\ee 
where the numerical values correspond to our universe. Stellar 
masses are roughly comparable to this benchmark scale, in our 
universe \citep{phil} and others \citep{adams}. Note that the 
mass scale $\starmass$ is equivalent to the Chandrasekhar mass 
\citep{chandra} with all of the numerical constants set to unity. 

Constraints on habitable planets can be divided into two conceptually
different categories. The first of class involves the stellar
properties associated with habitable planets. In addition to the need
for viable stellar structure solutions (working stars), we also
require that the host stars live long enough for biological evolution
to occur and have surface temperatures high enough to drive chemical
reactions. Note that the relevant time scales and energy levels depend
on the fine structure constant $\alpha$.  These stellar constraints
are considered in Section \ref{sec:stellarcon}.  The second class of
constraints concerns the relative ordering of the mass scales involved
in producing and maintaining habitable planets.  Some of these
considerations involve properties of the planets themselves, including
the requirement that the planets are massive enough to support a
biosphere and to retain an atmosphere, but not so massive as to become
degenerate. In addition, we require that planets are smaller than
their host stars and that stars are smaller than their host
galaxies. These issues are addressed in Section \ref{sec:ordering}.
The paper concludes in Section \ref{sec:conclude} with a summary of
the results and a discussion of their implications. In order for a 
universe to have working stars and habitable planets, the structure 
constants cannot vary by more then a few orders of magnitude from 
their measured values. 

\section{Constraints from Stellar Properties}
\label{sec:stellarcon} 

\subsection{Stellar Structure Solutions and the Existence of Stars}  
\label{sec:starstruct} 

To evaluate habitability constraints that depend on stellar
properties, we need a working model for stars in other universes (with
different values for the fundamental constants). Toward this end, we
use the semi-analytical model of \citep{adams}. This treatment solves
the standard equations of stellar structure
\citep{chandra,phil,kippenhahn,hansen}, but makes a number of
simplifying assumptions in order to obtain semi-analytic results:
First, the physical structure of the star is taken to be that of a
polytrope (with index denoted as $n$). Another simplification is that
only a single chain of nuclear reactions, characterized by a single
nuclear reaction rate, is considered.  The resulting model reproduces
stellar properties in our universe (including luminosity $L_\ast$,
temperature $T_\ast$, and radius $R_\ast$) to within a factor of
$\sim2$, while the stellar mass $M_\ast$ varies by a factor of
$\sim1000$ and the luminosity varies by a factor of $\sim10^{10}$.
Although approximate, the resulting stellar structure model is robust
enough to provide solutions across a parameter space where $\alpha$
and $\alpha_G$ vary by ten orders of magnitude in either direction
from their values in our universe.

In this model, the central temperature of the star is given by 
the solution to the equation 
\be 
\thetacen I(\thetacen) \tcent^3 = 
{ (4 \pi)^3 a c \over 3 \beta \kappa_0 \conlum } 
\left( {M_\ast \over \mzero} \right)^4 
\left( {G \over (n + 1) \rgas } \right)^7 \, , 
\label{tsolution} 
\ee 
where $\thetacen$ is related to the central temperature $\tcent$ 
through the expression 
\be
\thetacen = \left( {E_G \over 4 k \tcent} \right)^{1/3} 
\qquad {\rm where} \qquad E_G = \pi^2 \alpha^2 \mpro c^2 \,, 
\ee
and where the Gamow energy $E_G \approx 493$ keV for hydrogen fusion
in our universe.\footnote[2]{The expression for the Gamow energy $E_G$ 
assumes equal mass reacting particles with unit charge.}  In equation
(\ref{tsolution}), $\mzero$ and $\beta$ are dimensionless parameters
of order unity; they are determined by the mass and luminosity
integrals over the structure of the star, as characterized by the
polytropic index $n$. The parameter $\rgas$ is the gas constant that
appears in the ideal gas law, and $\kappa_0$ is the benchmark value of
the stellar opacity. Finally, the composite parameter $\conlum$
determines the nuclear reaction rate. Note that $\conlum$ depends on
the mass of the reacting particles, their charges, the mean energy
generated per nuclear reaction, and the ratio of the overall
coefficient of the nuclear cross section to the fine structure
constant (see \citep{adams}). For the sake of definiteness, we assume
here that $\conlum$ is constant, while $\alpha$ and $\alpha_G$ are
allowed to vary; additional variations should be considered in future
work.

The function $I(\thetacen)$ is defined by the integral of the
luminosity density over the stellar volume, i.e., 
\be
I(\thetacen) = \int_0^{\xi_\ast} \xi^2 d\xi f^{2n} \Theta^2 
\exp\left[ - 3 \Theta \right] \,,
\label{iintegral} 
\ee
where $\Theta = \thetacen f^{-1/3}(\xi)$, and 
where $f(\xi)$ is the solution to the Lane-Emden equation 
\citep{phil,chandra,kippenhahn,hansen}. The function $I(\thetacen)$ 
can be approximated by a fitting function of the form 
\be
\thetacen I(\thetacen) = B \thetacen^b \exp [-3 \thetacen] \,.
\ee
For polytropic index $n$ = 3/2, the fitting parameters have values $B$
= 0.833 and $b$ = 2.30.\footnote[2]{We can also use the full numerical  
solution to the integral in equation (\ref{iintegral}), but the
results are the same.}  The solution for the central temperature can
be written in the alternate form 
\be 
I(\thetacen) \thetacen^{-8} = {2^{12} \pi^5 \over 45} 
{1 \over \beta \kappa_0 \conlum E_G^3 \hbar^3 c^2} 
\left( {M_\ast \over \mzero} \right)^4 
\left( {G \mbar \over n+1 } \right)^7 \, ,
\label{tcsolution} 
\ee 
where $\mbar$ is the mean mass of the particles that make up the star. 
Note that the right hand side of the equation is dimensionless. For
the typical parameter values in our universe, the right hand side of
this equation has a value of approximately $10^{-9}$.

With the central temperature $\thetacen$ determined through equation
(\ref{tcsolution}), the equations of stellar structure specify the
remaining the properties of the star. The stellar radius $R_\ast$ is
given by 
\be
R_\ast = {G M_\ast \mbar \over k \tcent} 
{\xi_\ast \over (n+1) \mzero} \, ,
\label{rstar} 
\ee 
where $\xi_\ast$ is the dimensionless radius of the star (and is of 
order unity). The stellar luminosity $L_\ast$ takes the form  
\be
L_\ast = {16 \pi^4 \over 15} {1 \over \hbar^3 c^2 \beta \kappa_0 \thetacen} 
\left( {M_\ast \over \mzero} \right)^3 \left( 
{G \mbar \over n + 1 } \right)^4 \, . 
\label{lumstar} 
\ee 
The photospheric temperature $T_\ast$ of the star is then determined from 
the outer boundary condition so that 
\be
T_\ast = \left( {L_\ast \over 4 \pi R_\ast^2 \sigmasb} \right)^{1/4} \,,
\label{tphoto} 
\ee   
where $\sigmasb$ is the Stefan-Boltzmann constant. 

\subsection{Minimum Stellar Temperatures}  
\label{sec:startemp} 

The surface temperature $\tplan$ of a planet is determined by balancing the 
heating from the central star and the radiated heat of the planet. Using 
the simplest treatment we obtain 
\be
\sigmasb \tplan^4 = f_T {L_\ast \over 16 \pi d^2} \,, 
\ee
where $f_T$ is an efficiency factor that takes into account both the
radiation reflected away from the planet and the heat retained by the
atmosphere. Here we assume that the planetary orbit is circular with
radius $d$. The temperature $T_{\rm B}$ required to drive chemical
reactions, and hence support biological operations, is derived in 
Section \ref{sec:atmosphere}; following equation (\ref{biotemp}), 
we write this temperature in the form $kT_{\rm B}$ =
$\epsilon\alpha^2\melect c^2$, where the efficiency factor 
$\epsilon\sim10^{-3}$ for terrestrial chemical reactions. 
If we then require that the planet is warm enough to support life,
$\tplan \ge T_{\rm B}$, and use the fact that the orbital radius must
exceed the stellar radius, $d \ge R_\ast$, we obtain the constraint 
\be
{L_\ast \over R_\ast^2} \simgreat {16 \pi \sigmasb \over f_T} 
\left( {\epsilon \alpha^2 \melect c^2 \over k} \right)^4 \,.
\label{ratbound} 
\ee
If we scale to values in our universe, where $T_{\rm B}\sim300$ K, 
we obtain the requirement 
\be
{L_\ast \over R_\ast^2} \simgreat 2.3 \times 10^7 
{\rm erg}\,{\rm sec}^{-1}\,{\rm cm}^{-2}\, 
\left( {\alpha \over \alpha_0} \right)^8 \,. 
\label{habit} 
\ee 

Given the solutions for the stellar luminosity and stellar radius 
found in the previous subsection, the ratio $L_\ast/R_\ast^2$ 
is given by 
\be
{L_\ast \over R_\ast^2} = 
{16 \pi^4 \over 15} {1 \over \hbar^3 c^2 \beta \kappa_0 \thetacen} 
\left( {M_\ast \over \mzero} \right) 
\left( {G \mbar \over n + 1 } \right)^2  
\left( {k \tcent \over \xi_\ast} \right)^2
\ee
The right hand side of this equation is an increasing function of
stellar mass. In order to satisfy the constraint for habitability, we
require that the ratio $L_\ast/R_\ast^2$ is larger than the lower
bound given in equation (\ref{ratbound}). A necessary condition is
thus that the maximum value of the ratio $L_\ast/R_\ast^2$ must be 
larger than this lower bound, which implies that the following
constraint must be met 
\be
{\pi^4 \over 15} {1 \over \hbar^3 c^2 \beta \kappa_0 \thetacen^7} 
\left( {M_\ast \over \mzero} \right)_{\rm max}  
\left( {G \mbar \over n + 1 } \right)^2  
\left( {E_G \over \xi_\ast} \right)^2 > 
{16 \pi \sigmasb \over f_T} 
\left( {\epsilon \alpha^2 \melect c^2 \over k} \right)^4 \,.
\label{templimit} 
\ee

To move forward, we need to determine the maximum stellar mass for a
given set of fundamental constants. As the mass of a star increases,
the fraction of its internal pressure that is provided by radiation
pressure (instead of gas pressure) increases. Let $\bcon$ denote the
fraction of the pressure provided by the ideal gas law, so that
$(1-\bcon)$ is the fraction provided by radiation. The star becomes
unstable when the radiation pressure dominates \citep{phil}; here 
we equate the maximum stellar mass with that for which the fraction 
has a critical value $\bcon\approx1/2$. The maximum stellar
mass is then given by the expression
\be
M_{\ast {\rm max}} =  
\left( {18 \sqrt{5} \over \pi^{3/2} } \right) 
\left( {1 - \bcon \over \bcon^4} \right)^{1/2}  
\left( {\mpro \over \mbar} \right)^2 \, \starmass \approx 50 \starmass \,, 
\label{maxmass}
\ee
where $\starmass$ is the fundamental stellar mass scale defined by 
equation (\ref{mscale}). 

Next we want to substitute the maximum mass scale (equation
[\ref{maxmass}]) into the stellar structure solution for the central
temperature (from equation [\ref{tcsolution}]), 
\be 
I(\thetacen) \thetacen^{-8} = {2^{12} \pi^5 \over 45} 
{\hbar^3 c^4 \over \beta \kappa_0 \conlum E_G^3} 
{(50)^4 \over \mzero^4} 
\left( {G \mbar^7 \over (n + 1)^7 } \right) \mpro^{-8} \,,
\ee 
as well as the constraint on the planetary surface temperature 
(from equation [\ref{templimit}]), 
\be
{\pi^3 \over 15} {1 \over \hbar \beta \kappa_0 \thetacen^7} 
\left({G \over \hbar c} \right)^{1/2} 
\left( {50 \over \mzero} \right) 
\left( {\mbar \over \mpro (n + 1)} \right)^2
\left( {E_G \over 4 \xi_\ast} \right)^2 > 
{\sigmasb \over f} 
\left( {\epsilon \alpha^2 \melect c^2 \over k} \right)^4 \,.
\ee

Now we can simplify the expressions further. Let $\mbar=\mpro$, 
$n=3/2$, and use the definition of $E_G$, so that the central 
temperature is given by 
\be 
I(\thetacen) \thetacen^{-8} = {2^{23} \pi^5 \over 9} 
{\hbar^3 c^4 \over \beta \kappa_0 \conlum E_G^3} 
{G \over \mzero^4} \mpro^{-1} \, 
\ee 
and the constraint takes the form 
\be
{\pi^3 \over 30} 
{E_G^2 \over \hbar \kappa_0 \thetacen^7} 
\left( {G \over \hbar c } \right)^{1/2} 
\left( {1 \over \beta \mzero \xi_\ast^2} \right) > 
{\sigmasb \over f} 
\left( {\epsilon \alpha^2 \melect c^2 \over k} \right)^4 \,.
\ee
This constraint on the fundamental constants is required for 
stars to have surface temperatures hot enough to support 
viable biospheres. 

\subsection{Minimum Stellar Lifetimes} 
\label{sec:lifetime} 

For a universe to be habitable, at least some of its stars must live
long enough for biological evolution to take place. Because the lowest
mass stars live the longest, so we can derive a constraint on the
fundamental parameters by considering the smallest possible stars.
Previous work \citep{adams,phil} shows that the minimum mass necessary
to sustain nuclear fusion can be written in the form 
\be
M_{\ast{\rm min}} = 6 (3\pi)^{1/2} \left({4\over5}\right)^{3/4} 
\left({kT_{\rm nuc} \over \melect c^2}\right)^{3/4} \starmass\,,
\label{minmass} 
\ee
where $\starmass$ is the fundamental stellar mass scale given by equation
(\ref{mscale}). If we invert equation (\ref{minmass}), it determines
the maximum temperature $T_{\rm nuc}$ that can be obtained with a star
of a given mass, where this temperature is an increasing function of
stellar mass. By using the minimum stellar mass from equation
(\ref{minmass}) to specify the mass in equation (\ref{tcsolution}), we
obtain the minimum value of the stellar ignition temperature. This
central temperature, or equivalently the value of $\thetacen$, 
is determined by solving the following equation  
\be
\thetacen I(\thetacen) = \left( {2^{23} \pi^7 3^4 \over 5^{11} } \right) 
\left( {\hbar^3 \over c^2} \right) \left( {1 \over \beta \mzero^4} \right) 
\left( {1 \over \mpro \melect^3} \right) \left( {G \over \kappa_0 \conlum} \right) \, . 
\label{iprofile}
\ee
The parameters on the right hand side of the equation have been
grouped to include pure numbers, constants that set units,
dimensionless quantities from the polytropic solution, particle
masses, and finally the stellar parameters that depend on the
fundamental constants. In the context of this paper, these latter
quantities can vary from universe to universe. Note that this
expression has been simplified by setting $\mbar=\mion=\mpro$ and by
using the polytropic index $n$ = 3/2. Allowing other choices for the
particle masses and the polytropic index leads to the right hand side
of equation (\ref{iprofile}) changing by a factor of order unity,
whereas we vary $(G/\kappa_0)$ (equivalently, $\alpha$ and $\alpha_G$)
by many orders of magnitude.

The stellar lifetime $t_\ast$ can be written in the form 
\be
t_\ast = {f_{\rm c} \effish M_\ast c^2 \over L_\ast} = 
{9375 \over 256 \pi^4} f_{\rm c} \effish \hbar^3 c^4 \beta \mu_0^3 
\kappa_0 \thetacen M_\ast^{-2} 
\left( G \mbar \right)^{-4} \, , 
\ee
where $f_{\rm c}$ is the fraction of the stellar fuel that is available for
fusion and $\effish$ is the efficiency of nuclear fuel conversion
(where $\effish \approx 0.007$ in our universe). Solar-type stars have 
access to a fraction $f_{\rm c} \approx 0.1$ of their nuclear fuel during 
their main-sequence phase, whereas smaller stars have larger $f_{\rm c}$ 
\citep{mdwarf,al1997}.  

We want to compare the stellar lifetime to the time scale for 
atomic reactions, where this latter quantity is given by 
\be
t_{\rm A} = {\hbar \over \alpha^2 \melect c^2} \,. 
\ee
In our universe, this atomic time scale has the value 
$t_{\rm A}\sim2\times10^{-17}$ sec. In comparison, in order for
biological evolution to develop complex life forms (observers) on
Earth, the required time scale was of order 1 Gyr, which corresponds
to $\sim10^{33}$ ticks of the atomic clock.  Unfortunately, we
currently have a sample size of one for the specification of the time
required for biological evolution; we are thus left with enormous
uncertainty. Suppose, for example, that the time necessary for the
development of life has a wide distribution.  In this case, it could
be possible that (i) the probability of life originating within 1 Gyr
could be low, but that (ii) the minimum time required for life to
develop (anywhere) could sometimes be much less than 1 Gyr
\cite{carter1983,carter2008}. As a result, the fiducial time scale of
1 Gyr, while appropriate for life on Earth, does not represent a
definitive limit. Given these uncertainties, we consider a range of
values, but use $10^{33}$ atomic time scales as the center of the
allowed range (see below).

In general, the ratio of the stellar lifetime to the atomic 
time scale takes the form 
\be
{t_\ast \over t_{\rm A}} = 
{9375 \over 256 \pi^4} f_{\rm c} \effish \hbar^2 c^6 \beta \mu_0^3 
\kappa_0 \alpha^2 \melect \thetacen M_\ast^{-2} 
\left( G \mbar \right)^{-4} \, . 
\ee
In other universes, the largest possible value of this ratio, 
corresponding to the smallest, long-lived stars, is thus given by 
\be
\left( {t_\ast \over t_{\rm A}} \right)_{\rm max} = 
\left( {5^{13/2} \over 9 \pi^8 2^{10}} \right) 
\left( {c^3 \over \hbar} \right) 
\left( f_{\rm c} \effish \beta \mu_0^3 \right) 
\left( {\melect^{5/2} \mpro^{5/2} \over \mbar^4} \right)
\left( {\kappa_0 \over G \alpha} \right) 
\thetacen^{11/2} \,,
\label{timerat} 
\ee
where we have grouped the various factors as before.  Note that
equation (\ref{timerat}), as written, depends on the temperature
parameter $\thetacen$, which is specified via equation
(\ref{iprofile}). We can thus combine equations (\ref{iprofile})
and (\ref{timerat}) to solve for the ratio of time scales, and 
set it equal to the minimum required for life to develop (here 
we use $t_\ast/t_{\rm A} > 10^{33}$ as described above). 

\begin{figure}[tbp]
\centering 
\includegraphics[width=.90\textwidth,trim=0 150 0 150,clip]{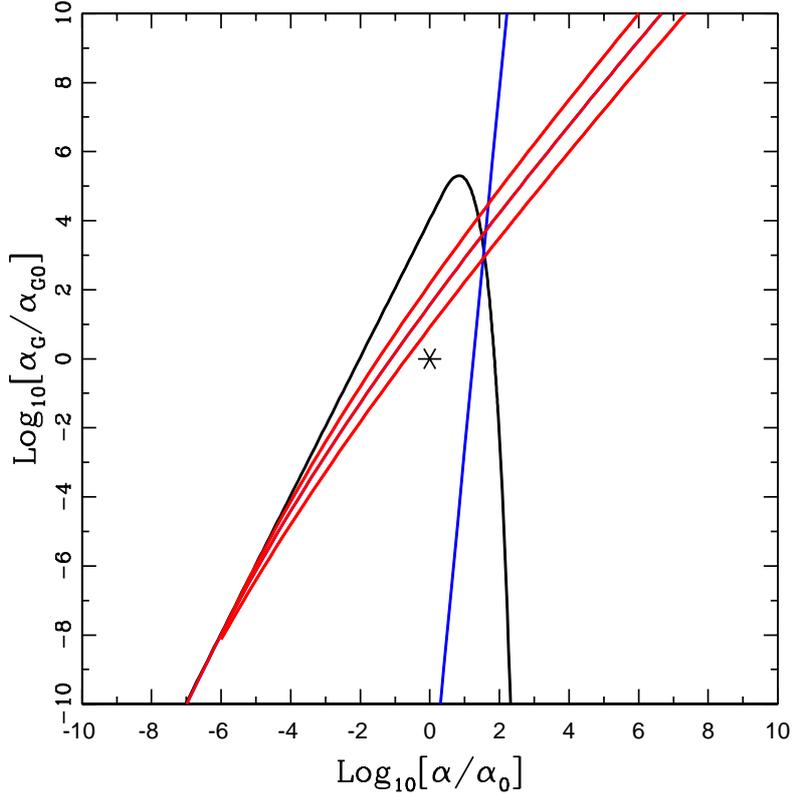}
\caption{\label{fig:starplane} Allowed plane of parameter space for 
varying structure constants $\alpha$ and $\alpha_G$, subject to
constraints on stellar properties. Both parameters are scaled to the
values in our universe. The region under the solid black curve
delineates the parameter space that allows stable long-lived stars to
exist (from \citep{adams}). In order to host habitable planets, stars
must have surface temperatures higher than the value required for
chemical reactions, where the viable parameter space falls to the
upper left of the solid blue line. Stars must also live long enough 
to allow for biological evolution. The viable parameter space falls 
to the lower right of the red curves, which are plotted for the 
required number of atomic time scales varying from $10^{32}$ (top) 
to $10^{34}$ (bottom). The star symbol denotes the position of our 
universe in the diagram. }
\end{figure} 

\subsection{Summary of Constraints from Stellar Structure} 
\label{sec:starsum}  

The results of this section are summarized in Figure
\ref{fig:starplane}, which shows the allowed plane of parameter space
for the structure constants $\alpha$ and $\alpha_G$.  The location of
our universe in the diagram is marked by the star symbol.  As outlined
below, the figure includes the constraints on parameter space
determined from considerations of stellar structure (compare with
Figure \ref{fig:planplane}, which shows the analogous constraints from
planetary considerations).

The first requirement is that working stars exist. The area below the
black curve represents the region for which long-lived stable stellar
configurations can sustain nuclear fusion.  Note that stars can fail
to exist for two conceptually different reasons: If the fine-structure
constant $\alpha$ is too large, then nuclear reactions are suppressed
and stars fail to generate nuclear power. When this condition occurs,
the minimum stellar mass (from equation [\ref{minmass}]) becomes
larger than the maximum stellar mass (from equation [\ref{maxmass}]).
On the other hand, if $\alpha$ is too small, then stable stellar
configurations cannot exist (equation [\ref{tcsolution}] has no
solution). Further, both of the aforementioned constraints depend on 
$\alpha_G$. For sufficiently strong gravity (large $\alpha_G$), the
range of $\alpha$ that supports working stars shrinks to zero.

We also require that the stellar photospheric temperature (equation
[\ref{tphoto}]) is larger than the temperature required for a working
biosphere (equation [\ref{biotemp}]). This condition (see Section
\ref{sec:startemp}) is marked by the steeply rising blue line in
Figure \ref{fig:starplane}. The viable regime of parameter space falls
to the (upper) left of the line.  To obtain this particular curve, we
require the surface temperature of the star to be larger than the
benchmark value $T_{\rm B}$ = 300 K $(\alpha/\alpha_0)^2$.  Although
we expect this scaling with $\alpha$ to hold, the exact value of the
coefficeint for $T_{\rm B}$ is not known, so that we should consider a
range of possible temperatures. In practice, however, the constraint
has such sensitive dependence on $T_{\rm B}$ that the effect of
including a range of values only serves to add width to the line shown
in Figure \ref{fig:starplane}.

Next we require that the stars live enough enough to allow for
biological evolution to take place. To invoke this constraint, we
start with the requirement that stars live for at least 1 Gyr in our
universe, where this constraint corresponds to $10^{33}$ atomic time
scales (allowing for the variation in chemical time scales due to 
varying $\alpha$). Since the number of required atomic time scales is
not known, we consider time scales both larger and smaller by an order
of magnitude. The resulting three curves are shown in red in Figure
\ref{fig:starplane}, where the allowed region of parameter space falls
below the curves. 

Each point of parameter space allows for a range of stellar masses.
To determine the constraints shown in Figure \ref{fig:starplane}, we
have used the minimum stellar mass to evaluate the lifetime constraint
and the maximum stellar mass to evaluate the temperature constraint.
One might worry that the region of allowed parameter space could be
smaller: The long-lived small stars might not have high enough surface
temperatures to maintain biospheres and/or the hot large stars might
not live long enough to allow for biological evolution. However,
numerical exploration shows that this complication does not reduce the
allowed region of parameter space.  On the right side of the plane,
large values of $\alpha$ lead to high mass stars becoming too cool;
the same (large) values of $\alpha$ allow the stellar lifetimes to be
long enough.  On the left side of the plane, small values of $\alpha$
lead to low-mass stars burning too quickly; these same (small) values
of $\alpha$ allow for sufficiently hot stellar photospheres.

The remaining region of allowed parameter space shown in Figure
\ref{fig:starplane} is relatively large. More specifically, if we fix
the gravitational constant $\alpha_G$ to the value in our universe,
the allowed range for the fine structure constant $\alpha$ spans about
an order of magnitude in either direction. The constraints from
stellar lifetimes and stellar temperatures are thus significant: If we
only require working stars, the range of $\alpha$ (black curve)
extends roughly two orders of magnitude in either direction.  The
allowed range in $\alpha$ becomes much wider for weaker gravity, but
disappears altogether if gravity is stronger by a factor of $\sim1000$
(compared to our universe). In addition, the allowed range of
$\alpha_G$ is not bounded from below: As gravity becomes weaker, stars
can still operate, but they require increasingly larger masses. At
some sufficiently small value of $\alpha_G$, we expect the required
large stellar masses to become problematic (see Section
\ref{sec:ordering}), but the issue is one of star formation (and
galactic considerations) rather than stellar structure.

\section{Constraints from Ordering of Mass Scales} 
\label{sec:ordering} 

\subsection{Maximum Planet Masses from Degeneracy}
\label{sec:degenerate} 

A planet must be supported primarily by electromagnetic forces. As a
result, the existence of planets requires that the electromagnetic
self-energy of a body exceeds the energy due to self-gravity
\citep{weisskopf}. The gravitational energy is given by the expression 
\be
E_{\rm g} = - f_n {G \mplan^2 \over \rplan} \,,
\ee
where $\mplan$ is the planet mass and $\rplan$ is the planet radius.
The dimensionless constant $f_n$ is of order unity and depends on the
density distribution within the planet. If we model the planet as a
polytrope, then $f_n=3/(5-n)$, where $n$ is the polytropic index
\citep{chandra}.  For an $n=1$ polytrope, which provides a reasonably
good model for large planets, we thus obtain $f_n=3/4$.  The
electromagnetic energy is given by the expression 
\be
E_{\rm em} = N {e^2 \over \ell} \,,
\ee
where $N$ is the number of atoms in the planet, $e$ is the charge, 
and $\ell$ is the effective distance between charges. On average, 
the distance $\ell$ is given in terms of the mean number density, 
\be
\ell = \langle n \rangle^{-1/3} 
\qquad {\rm where} \qquad 
\langle n \rangle = {3 N \over 4\pi \rplan^3} \,. 
\ee
Combining the above equations gives us an approximate expression 
for the electromagnetic energy 
\be
E_{\rm em} = N^{4/3} e^2 \left({3\over4\pi}\right)^{1/3} 
\rplan^{-1}\,. 
\ee 
In order for the electromagnetic self-energy to exceed the
gravitational energy, $E_{\rm em} > |E_{\rm g}|$, the following constraint
must be satified: 
\be
N^{4/3} e^2 \left({3\over4\pi}\right)^{1/3} > f_n G\mplan^2  \,. 
\ee
Next we assume that the planet is made of a single type of 
atom of mass $A \mpro$ so that 
\be
\mplan = N A \mpro \,. 
\ee
The constraint then simplifies to the form 
\be
N < \left({e^3 \over G^{3/2}A^3 \mpro^3}\right) 
\left({3\over4\pi f_n^3}\right)^{1/2} = 
\left({\alpha \over A^2 \alpha_G} \right)^{3/2}  
\left({3\over4\pi f_n^3}\right)^{1/2} \,. 
\label{ndegen} 
\ee

\medskip

\subsection{Minimum Planet Masses from Atmosphere Retention}
\label{sec:atmosphere} 

In this section we derive a lower limit on planetary masses by
requiring that the surface gravity is strong enough to hold on to an
atmosphere \citep{bartip,press1980,presslight}. In order for a planet
to support chemical reactions, its surface temperature cannot be too
small. Chemistry takes place on the scale of atoms, where the energy
levels are given by 
\be
E_n = - a {\melect c^2 \alpha^2 \over 2 n_r^2} \,, 
\ee
where $a=1$ for Hydrogen and $n_r$ is the radial quantum number. 
The constraint on the surface temperature of the planet can then 
be written in the form 
\be
k T > k T_{\rm B} \equiv \epsilon \, \melect c^2 \alpha^2 \,,
\label{biotemp} 
\ee
where the dimensionless parmaeter $\epsilon$ incorporates any
additional uncertainties. Based on terrestrial chemistry, we expect
$\epsilon \sim 0.001$. In order for air molecules to remain bound to
the planetary surface, so that the atmosphere does not evaporate
quickly, the temperature must be less than the gravitational 
binding energy, i.e.,  
\be
kT < {G \mplan \mu_{\rm a} \over \rplan} \,,
\ee
where $\mu_{\rm a}=A_{\rm a} \mpro$ is the mass of an air molecule. 
Combining the two constraints from above, we find 
\be
\epsilon \, \melect c^2 \alpha^2 < 
{G \mplan \mu_{\rm a} \over \rplan} \,. 
\label{atm} 
\ee
To go further we need to estimate the planetary radius. Here we 
assume that the atoms are close-packed and have radius given 
by the Bohr radius 
\be
a_0 = {\hbar \over \melect c \alpha} \,.
\ee
The number of atoms in the planet is then given by 
\be
N = {\rplan^3 \over a_0^3} \qquad {\rm or} \qquad 
\rplan = N^{1/3} a_0 = {N^{1/3} \hbar \over \melect c \alpha} \,.
\ee
Using this result in equation (\ref{atm}), we derive a 
lower bound on the number of atoms in the planet
\be
N > \left( {\epsilon \over A A_{\rm a}} \right)^{3/2} 
\left( {\alpha \over \alpha_G} \right)^{3/2} \,.
\label{lowerb} 
\ee

Note that if we combine this lower bound on $N$ with the upper bound
derived in the previous section, we obtain the combined constraint 
\be
\left( {\epsilon \over A A_{\rm a}} \right)^{3/2} 
\left( {\alpha \over \alpha_G} \right)^{3/2} < N < 
\left({\alpha \over A^2 \alpha_G} \right)^{3/2}  
\left({3\over4\pi f_n^3}\right)^{1/2} \,. 
\label{squeeze} 
\ee
Since both sides of the expression depend on the structure 
constants in the same way, this constraint reduces to the form 
\be
\epsilon < {A_{\rm a} \over A f_n} 
\left({3\over4\pi}\right)^{1/3} \,. 
\ee
We expect the right hand side to be of order unity, whereas 
$\epsilon \sim 0.001$, so that this constraint is generally satisfied.

\subsection{Minimum Planet Masses from Biosphere Complexity} 
\label{sec:biosphere} 

Another lower limit on planet masses arises from the requirement that
planets must be large enough to support a biosphere. In approximate
terms, life can be described as a physical process that requires
information, and a certain minimum amount of information must be
processed for a planet to support life \citep{dyson}. In order to
function, for example, a human being requires a minimum information of
roughly $Q_1 \approx 10^{23}$ bits, whereas the human species as a
whole requires approximately $Q_T \approx 10^{33}$ bits.  A fully
functioning biosphere is thus expected to have some minimum value
$Q_B$ (where we expect $Q_B > Q_T$).  Our own biosphere is estimated
to have a mass of 500 to 800 billion tons of carbon, which corresponds
to about $4\times10^{40}$ particles. We thus write the minimum size of 
a biosphere in the form 
\be
Q_B = f_{\rm B} 10^{40} \, {\rm bits}\,, 
\ee
where the dimensionless parameter $f_{\rm B}$ encapsulates the
uncertainties in this quantity. The mass of the planet should thus 
be large enough so that its information content far exceeds this 
benchmark value.\footnote[2]{Note that this number of particles 
corresponds to the information content of the entire structure of 
the biosphere. The blueprint required to reproduce the biosphere, 
as encoded in the DNA base-pairs in all living cells, would be 
significantly smaller.} As a result, the minimum number of particles  
$N_{\rm min}$ in a potentially habitable planet is expected to 
obey the ordering $N_{\rm min} \gg Q_B = f_{\rm B} 10^{40}$. 
For the sake of definiteness, we define 
\be
N_{\rm min} \equiv f_{\rm bio} 10^{40} \,,
\label{nummin} 
\ee
where the dimensionless parameter $f_{\rm bio}$ is much larger 
than unity. For example, if the minimum size of a habitable 
planet is comparable to Earth, then $f_{\rm bio}\sim10^{11}$. 
This requirement, in conjunction with equation (\ref{ndegen}), 
places a constraint on the structure constants,  
\be
{\alpha \over \alpha_G} > N_{\rm min}^{2/3} 
A^2 f_n \left({4\pi \over 3}\right)^{1/3} 
\approx 5 \times 10^{28} f_{\rm bio}^{2/3} \,, 
\ee
or, alternately, 
\be
{\alpha_G \over \alpha_{G0}} \simless 2.5 \times 10^{7}\,
{\alpha \over \alpha_0}\,f_{\rm bio}^{-2/3} \,. 
\label{biocon} 
\ee
Keep in mind that $f_{\rm bio}\gg1$.

\subsection{Considerations of Stellar and Galactic Masses} 
\label{sec:stargalaxyorder} 

In this section we consider the constraints that planets should be
smaller in mass than their parental stars, and that stars are smaller
in mass than their host galaxies. It is not known if this ordering of
mass scales is strictly necessary for a universe to be habitable. If
this ordering is violated, however, the planets (being more massive
than stars) are likely to undergo nuclear fusion and therby become
uninhabitable. Moreover, the formation of stars (within galaxies) and
the formation of planets (within the disks asoociated with forming
stars) would be problematic.

As shown previously \citep{adams,phil}, stars have typical mass scales
given by $\starmass=\mpro\alpha_G^{-3/2}$ (see also equation
[\ref{mscale}] and Section \ref{sec:starstruct}), whereas the above
considerations indicate that planets have mass scales
$M_{0P}=\mpro(\alpha/\alpha_G)^{3/2}$ (equation [\ref{squeeze}]; see
also \citep{bartip}). If we require that planets are less massive than
their host stars, we obtain a constraint of the form 
\be
\alpha \simless 1 \qquad {\rm or} \qquad 
{\alpha \over \alpha_0} \simless 137\,.
\label{alphaone} 
\ee 
In addition to being smaller in mass than its host star, the planet
must have enough constituent particles to support a sufficiently
complex biosphere (Section \ref{sec:biosphere}). The coupled
requirements that $N_P > N_{\rm min}$ (equations [\ref{nummin} --
\ref{biocon}]) and $\alpha < 1$ (equation [\ref{alphaone}]) implies 
an upper limit on the gravitational constant, i.e.,
\be 
N_P > N_{\rm min} \qquad {\rm or} \qquad 
{\alpha_G \over \alpha_{G0}} \simless 3.38 \times 10^9 
f_{\rm bio}^{-2/3} \,,
\label{gmax} 
\ee 
where the factor $f_{\rm bio} \sim 1$ if the planet has the same
number of particles as our biosphere and $f_{\rm bio} \sim 10^{11}$ if
the planet has the same number of particles as Earth. Taken together,
equations (\ref{alphaone}) and (\ref{gmax}) imply that the structure
constants $(\alpha,\alpha_G)$ cannot be larger than those in our
universe by more than a few orders of magnitude.

Another constraint is provided by the requirement that stars should be
smaller in mass than their host galaxies. With the opposite ordering
of mass scales, star formation would be difficult. Unfortunately, the
masses of galaxies in other possible universes could vary enormously.
Even within our universe, a number of different physical processes
play a role in determining the masses of galaxies. Nonethless, we can
derive an expression for galactic masses, analogous to equation
(\ref{mscale}) for stars, through considerations of galaxy formation. 

During the process of galaxy formation, the mass scale $M_{\rm eq}$
contained in the cosmological horizon at the time of matter domination
plays an important role \citep{gungott,reesost,whiterees}. In brief,
growth on all scales is suppressed earlier during the radiation
dominated era, but perturbations with masses $M<M_{\rm eq}$ begin to
grow after the epoch of matter domination. Perturbations on even
larger mass scales $M>M_{\rm eq}$ can grow later, but their
development is delayed, and their eventual growth could be compromised
if the universe contains enough dark energy \citep{tegmark}.  Although
perturbations on all scales $M<M_{\rm eq}$ grow after the epoch of
matter domination, several complications arise: The virialization time
is somewhat shorter for the smaller dark matter halos, so they reach
nonlinearity first; the smaller halos collide and merge in a complex
tree of interactions; finally, growth is suppressed for the smallest
mass scales below the baryonic Jeans mass. In our universe, for
example, these considerations conspire to produce a cummulative mass
distribution of galactic halos that increases (slowly) up to a mass
roughly comparable to $M_{\rm eq}$ (e.g., see Figure 3.2 from
\citep{loeb}). We can thus consider $M_{\rm eq}$ as one proxy for
galactic masses in other universes.

The temperature at the epoch of matter domination is given by 
\be
k T_{\rm eq} = \eta (\mpro c^2) {\Omega_{\rm M} \over \Omega_{\rm b}} \,,
\ee
where $\eta$ is the baryon-to-photon ratio, $\Omega_{\rm M}$ is the
relative energy density in dark matter, and $\Omega_{\rm M}$ is the
relative energy density in baryons (note that we have ignored the
contribution of neutrinos in this simple treatment). The mass scale of
the horizon at this epoch is thus given approximately by 
\be
M_{\rm eq} = \left( {5 \over \pi} \right)^{1/2} {3 \over 64\pi} 
\alpha_G^{-3/2} \mpro \left( {\mpro c^2 \over kT_{\rm eq}} \right)^2 
\approx {\starmass \over 64 \eta^2} 
\left( {\Omega_{\rm b} \over \Omega_{\rm M}} \right)^2 \,,
\ee
where $\starmass$ is the stellar mass scale from equation (\ref{mscale}). 
In our universe, the baryon-to-photon ratio $\eta \sim 10^{-9}$, so
that this mass scale is larger than the stellar mass scale by a factor
of about $10^{15}$. In any case, the mass scale $M_{\rm eq}$ scales
linearly with $\starmass$, where the coefficient depends on
cosmological parameters $(\eta,\Omega_{\rm M},\Omega_{\rm b})$, and
not on the structure constants $(\alpha,\alpha_G)$. As a result, the
mass scale $M_{\rm eq}$ will always be much greater than the stellar
scale $\starmass$ as long as $\eta \ll 1$, so that we do not obtain 
an additional constraint. 

Although galactic halos can in principle form with mass scales
$M{\sim}M_{\rm eq}$, and with even larger masses at later epochs, the
resulting structures will not always produce stars.  In order for a
galaxy to form stars, and thereby become habitable, the baryonic gas
must cool on sufficiently rapid time scales
\citep{gungott,whiterees,reesost}. The requirement that the gas
cooling time is comparable to the free-fall collapse time for
cosmological structures can be used to specify a characteristic mass
scale for galaxies in terms of the structure parameters \citep{reesost}. 
The result for $M_{\rm gal}$ can be written in the form 
\be
M_{\rm gal} = \alpha_G^{-2} \alpha^5 
\left( {\mpro \over \melect} \right)^{1/2} \mpro = \starmass 
\alpha_G^{-1/2} \alpha^5 \left( {\mpro \over \melect} \right)^{1/2} \,,
\label{galcool} 
\ee
where $\starmass$ is the characteristic stellar mass. For our
universe, the mass scale of equation (\ref{galcool}) is larger than
the stellar mass scale by a factor of $\sim 10^{10}$, which makes
$M_{\rm gal}$ comparable in mass to an ordinary galaxy. This mass
scale is often used as characteristic galactic mass \citep{bartip,burost}.  
However, one should keep in mind that galaxies span a wide range of
masses, from about $10^7$ to $10^{13}$ $M_\odot$, i.e., a range of a
factor of $f_{\rm gal}\sim1000$ on either side of the scale given by 
equation (\ref{galcool}).  The constraint that stars are smaller in
mass than their host galaxies thus takes the form
\be
\alpha_G \simless f_{\rm gal}^2 \alpha^{10} {\mpro\over\melect} 
\qquad {\rm or} \qquad 
{\alpha_G \over \alpha_{G0}} \simless f_{\rm gal}^2 
(1.33 \times 10^{20}) \left( {\alpha \over \alpha_0} \right)^{10} \,.
\label{coolconst}
\ee
At first glance, this constraint might not seem restrictive. However,
with the large exponent for $\alpha$, the fine structure constant can 
only decrease by a factor of $\sim100$ before the relevant parameter 
space starts to shrink. 

\begin{figure}[tbp]
\centering 
\includegraphics[width=.90\textwidth,trim=0 150 0 150,clip]{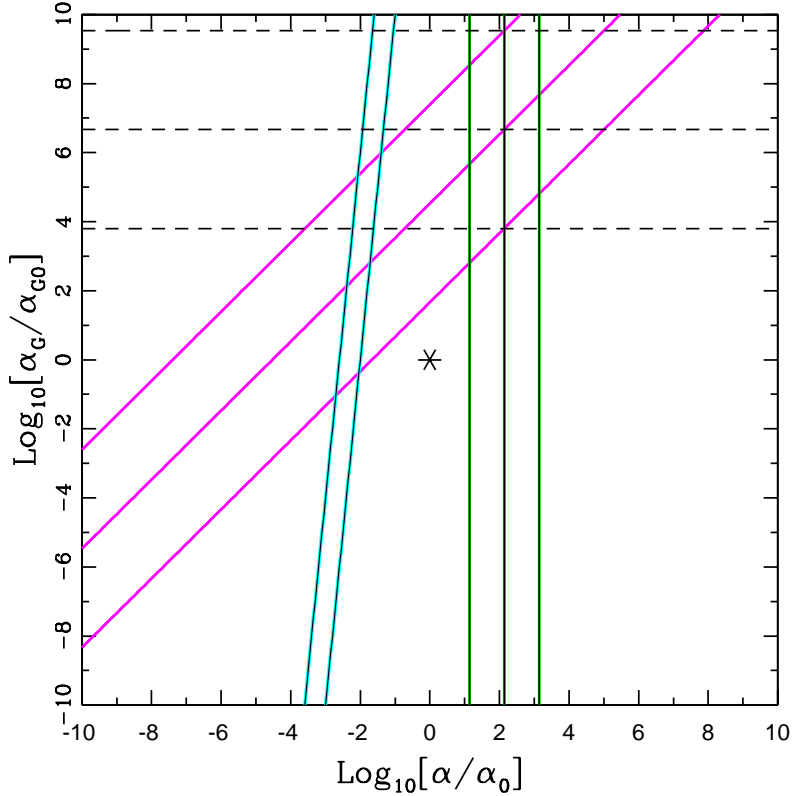}
\caption{\label{fig:planplane} Allowed plane of parameter space for 
the structure constants $\alpha$ and $\alpha_G$, where both parameters
are scaled to the values in our universe. The constraints shown here
arise from the ordering of astrophysical mass scales: Planets large
enough to support a complex biosphere and small enough to be
non-degenerate must fall below the diagonal magenta lines. For planets
to be less massive than their host stars, the values of $\alpha$ must
fall to the left of the vertical green lines.  For stars to be less
massive than their host galaxies, the value of $\alpha$ must lie to
the right of the cyan lines. For planets to be less massive than
their host stars and sufficiently complex to support a biosphere, the
value of $\alpha_G$ must fall below the black horizontal dashed lines.
Constraints are shown for a range of values (see text). The star
symbol denotes the position of our universe in the diagram. }
\end{figure}

\subsection{Summary of Constraints from Mass Scales} 
\label{sec:sumplanet} 

The constraints of this section are summarized in Figure
\ref{fig:planplane}, which shows the plane of parameter space for the
structure constants $\alpha$ and $\alpha_G$ (scaled to the values in
our universe). The location of our universe in the diagram is marked
by the star symbol.  This figure contains four types of boundaries to
the parameter space, as outlined below:

We first invoke the constraint that planets must be large enough in
mass to support a sufficiently complex biosphere and simultaneously
small enough to remain non-degenerate (see equation [\ref{biocon}]).
Figure \ref{fig:planplane} shows three choices for the minimum size of
a planet: The top diagonal magenta line corresponds to the requirement
that the planet contains the information content of our biosphere; a
viable planet must be at least as large as the biosphere it supports,
so this boundary in parameter space is overly generous.  The bottom
diagonal magenta line corresponds to the requirement that the planet
must be as large as the Earth (in particle number); since our planet
is clearly large enough to support a biosphere, this boundary is too
restrictive. The minimum planet size must fall between the previous
two choices.  Given the large parameter space, and the uncertainties,
a good benchmark value is provided by the geometric mean of the two
previously defined scales. This estimate implies that the planet must
be larger than a moon-sized body; the corresponding boundary is marked
by the central diagonal magenta line in the figure.

Planets must also be smaller than their host stars, where this
requirement implies that the fine structure constant $\alpha$ is
bounded from above (equation [\ref{alphaone}]). These constaints are
indicated by the vertical green lines for $\alpha<1/10,1$, and 10.
By combining the previous constraints, we obtain the horizontal dashed
lines near the top of Figure \ref{fig:planplane}: Here we require that
a potentially habitable planet must be less massive than its host star
(so that $\alpha\simless1$), but large enough to support a biosphere
(see equation [\ref{gmax}]).  The three lines shown in the figure
correspond to constraints requiring that the planet has the information
content of our biosphere (top line), the entire Earth (bottom line),
and the geometric mean (center line).

Similarly, stars must have smaller masses than their host galaxies.
Here we require galactic structures to cool on sufficiently short time
scales, so that they can form stars (see equation [\ref{coolconst}]).
The two diagonal cyan lines in Figure \ref{fig:planplane} define the
resulting allowed range of parameter space (which lies to the right of
the curves).  Results are shown for two choices of the dimensionless
parameter $f_{\rm gal}$ = 1 and 1000; these values correspond to
maximum galactic masses in our universe of $\sim10^{10}\starmass$ and
$\sim10^{13}\starmass$, respectively.

The surviving region of parameter space in the $\alpha$-$\alpha_G$
plane is relatively large: The width of the region corresponds to a
range in $\alpha$ spanning 4 to 7 orders of magnitude. For the portion
of parameter space shown in Figure \ref{fig:planplane}, the height of
the allowed region corresponds to a range in gravitational constant,
equivalently $\alpha_G$, spanning 8 to 10 orders of magnitude --- but
note that no lower limit on $\alpha_G$ exists. This allowed region of
parameter space is ``large'' in the sense that both structure
parameters can vary by many orders of magnitude. Nonetheless, several
caveats should be kept in mind: Although the parameter space is
presented here in terms of the quantities $\log_{10}[\alpha]$ and
$\log_{10}[\alpha_G]$, it is not known if the logarithm of
$(\alpha,\alpha_G)$ represents the proper weighting. In fact, the
underlying probability distributions for these structure parameters
remain unknown.  These (unspecified) probability distributions must
be convolved with the allowed region of parameter space in order to
make a full assessment of the likelihood of obtaining habitable
systems.

\section{Conclusions} 
\label{sec:conclude} 

\subsection{Summary of Results} 
\label{sec:results} 

\begin{figure}[tbp]
\centering 
\includegraphics[width=.90\textwidth,trim=0 150 0 150,clip]{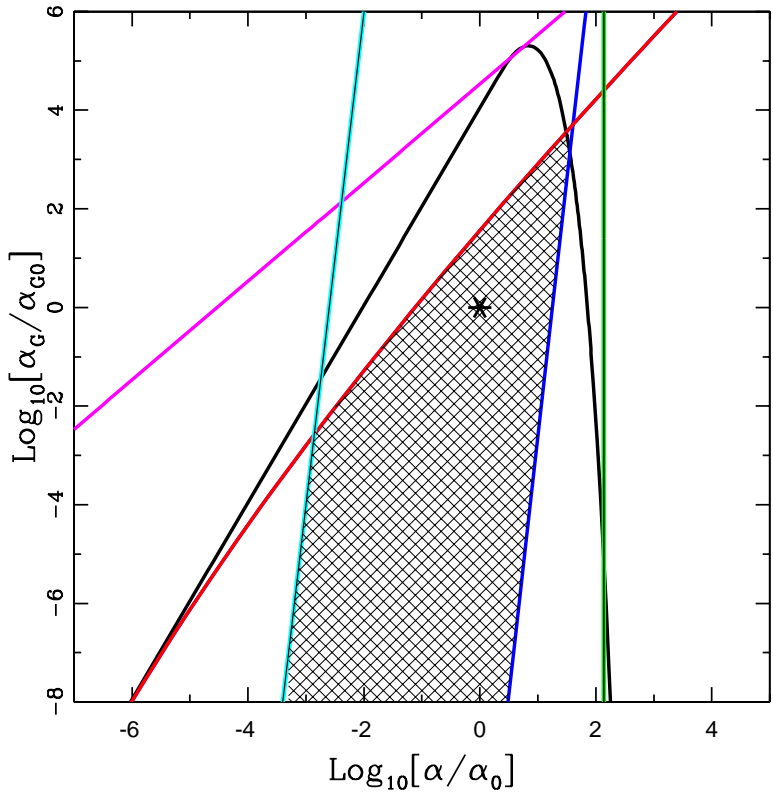}
\caption{\label{fig:compplane} Allowed plane of parameter space for 
the structure constants $\alpha$ and $\alpha_G$. The shaded region
delineates the portion of the plane that remains after enforcing the
constraints from this paper. The black curve shows the requirement
that stable stellar configurations exist. The blue curve shows the
requirement that the stellar temperature is high enough to allow
habitable planets. The red curve shows the constraint that stars live
long enough for biological evolution to occur ($10^{33}$ atomic time
scales). For planets to be smaller than stars, the fine structure
constant $\alpha$ must lie to the left of the vertical green line. For
stars to be smaller than their host galaxies, $\alpha$ must fall to
the right of the cyan curve. For planets to carry enough information
content to support a biosphere, and remain non-degenerate, the
parameters must fall below the diagonal magenta line. }
\end{figure}

This paper has developed a number of constraints that limit the
parameter space for universes that support both working stars and
habitable planets. Previous work shows that stars can exist over a
reasonably wide portion of the $\alpha$-$\alpha_G$ plane of parameters
\citep{adams}. For small values of $\alpha$, stars fail to exist
because of the absence of stable nuclear burning configurations; for
large values of $\alpha$, stars fail to exist because the allowed
range of stellar masses shrinks to zero. This work presents a number
of additional constraints that reduce the viable region of the
$\alpha$-$\alpha_G$ plane, including considerations of stellar
structure (Section \ref{sec:stellarcon} and Figure \ref{fig:starplane}), 
and planetary properties (Section \ref{sec:ordering} and Figure
\ref{fig:planplane}).  Using characteristic values to evaluate each
constraint, Figure \ref{fig:compplane} shows the remaining portion of
the $\alpha$-$\alpha_G$ parameter space. The area below the heavy
black curve represents the region for which working stars can exist.  
The other bounds can be summarized as follows: 

The requirement that stars have a sufficiently long life span can be
measured by the ratio of the stellar lifetime to the chemical (atomic)
time scale. Biological evolution requires a minimum number of ticks of
this atomic clock. Although the exact value remains unknown, we use
$10^{33}$ as a working estimate, which correponds to a time of $\sim1$
Gyr in our universe; we also use the smallest (longest-lived) stars to
evaluate the constraint.  This requirement reduces the the allowed
range of parameter space by eliminating large values of $\alpha_G$ and
small values of $\alpha$; the allowed region falls below the red curve
in Figure \ref{fig:compplane}.

We also enforce the constraint that planets can maintain a habitable
temperature. This condition is equivalent to requiring that the
stellar surface temperature is higher than the temperature necessary
to drive chemical reactions, i.e., the habitable zone must lie outside
the star.  Here we use the largest (hottest) stars to evaluate the
bound. This constraint reduces the allowed parameter space by
eliminating large values of $\alpha$. The constraint varies slowly
with the strength of gravity and is more stringent for smaller values
of $\alpha_G$; the allowed region of parameter space falls above (to
the left of) the blue curve in Figure \ref{fig:compplane}.

Planets must be large enough in mass to hold onto an atmosphere and
small enough in mass to remain non-degenerate. We have shown that
these requirements scale with the fundamental constants in the same
way and are generally satisfied. In addition, planets must be large
enough, in particle number, to support a biosphere of sufficient
complexity.  If we require the biosphere to be as complex as that of
our own, then the corresponding constraint is less restrictive than
the requirement of working stars. Even if we require that the
biosphere -- and hence the planet -- has the complexity of earth, very
little of the allowed parameter space is eliminated. The geometric
mean of these extremes in shown as the diagonal magenta line in Figure
\ref{fig:compplane}. Although this consideration reduces the allowed
parameter space, the need for sufficently long stellar lifetimes
provides a roughly parallel but stronger constraint.

The requirement that stars are more massive than planets implies an
upper limit on the fine structure constant $\alpha \simless 1$
(equivalently, $\alpha/\alpha_0\simless137$), as marked by the
vertical green line in Figure \ref{fig:compplane}. This constraint is
comparable to that required for stars to exist: If $\alpha$ is too
large, then electrostatic repulsion effectively shuts down nuclear
reactions in stars. The requirement of a sufficiently high
photospheric temperature (blue curve) provides an even stronger limit.
We thus have three independent reasons to disfavor universes with
large values of the fine structure constant. All three considerations
require that $\alpha/\alpha_0\simless100$ or $\alpha\simless1$.

Extremely weak gravity, corresponding to small values of $\alpha_G$,
leads to extremely massive stars, essentially because stellar masses
scale like $\starmass\sim\alpha_G^{-3/2}$. Additional constraints are
provided by the requirement that stars cannot be too massive. In order
for galaxies to form stars, gas must cool efficiently, and this
requirement defines a characteristic mass scale $M_{\rm gal}$ for
operational galaxies (equation [\ref{galcool}]). If we take the
maximum galactic mass to be $1000M_{\rm gal}$, the constraint that
stars are less massive than their host galaxies is delineated by the
cyan curve in Figure \ref{fig:compplane}. 

The combined constraints outlined above also provide global bounds on
the structure parameters $\alpha$ and $\alpha_G$.  Specifically, the
allowed region shown in Figure \ref{fig:compplane} indicates that
$\alpha/\alpha_0\simless37$ and $\alpha_G/\alpha_{G0}\simless3700$.  
A related quantity of interest is an upper bound on the ratio
$\alpha_G/\alpha$ of the structure parameters. For the viable
parameter space shown in Figure \ref{fig:compplane}, this ratio 
has a maximum value given by 
\be
{\alpha_G/\alpha_{G0} \over \alpha/\alpha_0} \simless 130 
\qquad {\rm or} \qquad 
{\alpha_G \over \alpha} \simless 10^{-34}\,.
\ee
This result shows that any universe with working stars and habitable
planets must have a large hierarchy between the strengths of its 
forces, i.e., the values of $\alpha$ and $\alpha_G$.

The bounds on the structure parameters described above were derived
from a numerical evalution of the coupled constraints of this paper
(see also \citep{reessix,rees1972}).  For completeness, we present a
simpler analytic approach to this problem in Appendix \ref{sec:appendix}.  
The resulting bounds are somewhat less constraining than those found
above, but they can be expressed in terms of simple analytic
expressions. These bounds indicate that $\alpha/\alpha_0\simless574$,
$\alpha_G/\alpha_{G0}\simless2\times10^6$, and
$\alpha_G/\alpha\simless2\times10^{-31}$.

\subsection{Discussion} 
\label{sec:discuss} 

Even with all of the constraints considered in this paper, the region
of allowed parameter space for the structure constants is still
relatively large. Specifically, if we work in terms of the
$\alpha$-$\alpha_G$ plane using logarithmic units and the range shown
in Figures \ref{fig:starplane} and \ref{fig:planplane}, then the
region that allows for working stars corresponds to about one fourth
of the original plane \citep{adams}.  The additional constraints
considered in this paper reduce the allowed region of the plane by
another factor of $\sim$two. Keep in mind, however, that this work
only delineates the region of the plane that allows for both working
stars and habitable planets (using a particular set of constraints).
Whether or not this region is ``small'' or ``large'', or if it
provides evidence for or against fine-tuning, depends on unknown
quantities: In particular, the proper measure of the parameter space
requires knowledge of the underlying probability distribution from
which universes select their values of the structure constants, in
this case $\alpha$ and $\alpha_G$. Nonetheless, the constants can vary
by significant factors (a few orders of magnitude) and still allow a
universe to remain viable.  We stress that additional constraints
would lead to further reduction of the allowed parameter space: For a
sufficiently large and restrictive set of constraints, the allowed
parameter space must collapse to the neighborhood of the point
representing our universe.

As shown in Figure \ref{fig:compplane}, this paper presents three
upper bounds on $\alpha$: planets must be smaller than stars (green
line), stars must burn nuclear fuel (black curve), and photospheres
must be hot enough (blue curve). Two additional constraints can also
be considered: First, in order for chemistry to operate properly, the
fine structure constant must be bounded from above. If $\alpha$ is too
large, then the innermost electrons in atoms become relativistic and
are subject to capture. As a rough approximation, in order to get $N$
different chemical species, one needs $\alpha < 1/N$; otherwise, the
universe in question would lose most of its periodic table
\cite{reesemail}. Second, in order for the proton to exist an an
allowed bound state of quarks, the fine structure constant is bounded
from above according to $\alpha<(m_{\rm d}-m_{\rm u})$/141MeV
$\approx$ 1/56, where $m_{\rm u}$ and $m_{\rm d}$ are the quark masses
\citep{lawrence,hogan}.  These considerations thus provide two more
reasons to disfavor universes with large $\alpha$.  In other words,
the fine structure constant is confined to the regime $\alpha \ll 1$,
so that electromagnetism must be perturbative.

Another interesting result emerges from the considerations of this
paper: The strongest limits on the existence of habitable planets
arise from constraints on stellar properties (in particular,
sufficiently long lifetimes and hot surface temperatures) rather than
constraints on the properties of planets themselves. The bounds
arising from planetary considerations, and ordering of mass scales,
are roughly parallel to those arising from stellar considerations, but
are somewhat weaker (see Figure \ref{fig:compplane}).  This finding
suggests that stars are the key element in determining the potential
habitability of a universe.

On one hand, this paper extends previous constraints on the range of
allowed values for the fundamental constants. On the other hand, the
treatment is specific to the question of planetary habitability and
the accompanying constraints on stellar properties. In the future,
this work should be extended in several directions: Although stars and
planets can exist within the range of parameter space found here, we
have not yet shown that such bodies are readily formed in these
alternate universes. Small values of $\alpha$ lead to inefficient
cooling, which can inhibit star formation; small values of $\alpha_G$
lead to large stellar masses, which also inhibit star formation unless
galaxies are correspondingly larger. Although we have included simple
considerations of galactic cooling (equation [\ref{galcool}]), future
work should place these formation constraints within a broader
astrophysical context.

The results presented here are obtained using relatively simple models
for stellar structure, as well as for planetary and galactic
considerations.  A more detailed treatment should be carried out in
the future. The current stellar structure calculations are also
limited to the hydrogen burning phase, i.e., we consider the fusion of
only one nuclear species. The production of heavy elements, including
carbon and oxygen, place additional constraints on the fundamental
constants, and thereby narrow the allowed range of parameters.

Another line of inquiry is to consider additional possible variations
of the fundamental constants. This paper has focused on $\alpha$ and
$\alpha_G$, but the strength of the nuclear forces, the masses of the
fundamental particles, and/or the cosmological parameters could also
vary in other universes \cite{reessix,hogan,tegmark,schell,weakless}.
For example, if the strong force were sufficiently weak, no bound
nuclei would exist, and no combination of the other constants would
allow for long lived, stable, nuclear burning stars.  On the other
hand, in a complete theory --- not yet available --- variations in the
fundamental parameters could be coupled, so that changes in one
constant are not independent from changes in another. In the context
of this paper, this latter constraint would imply that the allowed
parameter space traces out a particular curve through the
$\alpha$-$\alpha_G$ plane shown in the figures. As one example,
equation (54) of Ref. \citep{carr} suggests that $\alpha^{-1} \sim
\log \alpha_G^{-1}$, which would imply a nearly vertical path through
the plane (see also \citep{page}). Notice also that the strengths of
the weak and strong forces, and hence the composite nuclear parameter
$\conlum$, would also vary in such a coupled theory.  In any case,
however, the region of parameter space that allows for viable
universes does not seem to be prohibitively small, and much more work
must be done to delineate its boundaries.

\acknowledgments

We would like to thank Juliette Becker, Tony Bloch, Kate Coppess, Gus
Evrard, Evan Grohs, Dragan Huterer, Minhyun Kay, Greg Laughlin, James
Wells, and Coco Zhang for useful discussions. This work was supported
by JT Foundation grant ID55112 and by the University of Michigan.

\appendix
\section{Global Bounds on the Structure Constants} 
\label{sec:appendix}  

This Appendix derives upper bounds on the fine structure constant
$\alpha$, the gravitational constant $G$ (equivalently $\alpha_G$),
and on the ratio of force strengths $\alpha_G/\alpha$. The results
from the main text (see Figure \ref{fig:compplane}) follow from
numerical evaluation of all of the coupled constraints on the
parameter space. The bounds presented here are derived from simple
analytic considerations, but are weaker. 

\subsection{Definitions and Dimensionless Constraints} 
\label{sec:define} 

We start by writing all of the constraints in dimensionless form. 
The stellar mass can be written in terms of the fundamental stellar
mass scale $\starmass$ so that
\be
M_\ast = X \starmass = X \alpha_G^{-3/2} \mpro \,. 
\ee
The fine structure constant and the gravitational constant are written 
in terms of dimensionless factors according to 
\be
\alpha_{\rm univ} \equiv a \, \alpha
\qquad {\rm and} \qquad G_{\rm univ} \equiv g \, G \,,
\ee
where the un-subscripted quantities correspond to the values in our 
universe. Since we keep the particle masses constant, the second 
expression is equivalent to $\alpha_{G{\rm univ}} = g \alpha_G$.

The equation that sets the central temperature of the star necessary
for a long-lived stable configuration can then be written in the form
\be
I(\thetacen) \thetacen^{-8} = A X^4 g a^{-8} \,,
\label{starstruck} 
\ee
where the dimensionless constant $A$ is given by 
\be
A = \left( {2^{19} \pi^5 \over 9 \cdot 5^8} \right) 
\left( {1 \over \beta \mzero^4} \right) 
\left( {\hbar^3 c^4 \over E_G^3 \mpro} \right) 
\left( {G \over \kappa_0 \conlum} \right) 
\approx 5.23 \times 10^{-9} \,. 
\label{aconstant} 
\ee

The condition that stars have a minimum temperature can be written 
\be
B X g^{1/2} > a^6 \thetacen^7 \,,
\label{temperature} 
\ee 
where the dimensionless constant $B$ is given by 
\be
B = \left( {\pi \over 25} \right) 
\left( {1 \over \beta \mzero \xi_\ast^2} \right) 
\left( {E_G^2 \over \kappa_0} \right) 
\left( {G \over \hbar c } \right)^{1/2} 
{(\hbar c)^2 \over \left( \epsilon \alpha^2 \melect c^2 \right)^4} 
\approx 5.70 \times 10^{10} \,.
\label{bconstant} 
\ee
The condition that stars have a sufficiently long lifetime 
takes the form 
\be
C a^4 \thetacen > g X^2 \,,
\label{lifetime} 
\ee
where the constant $C$ is given by 
\be
C = \left( {9375 \over 256 \pi^4} \right) 
\left( f_{\rm c} \effish \beta \mzero^3 \right) 
\left( {\melect c^3 \over \hbar} \right) 
\left( {\kappa_0 \alpha^2 \over G} \right) 
{1 \over N_{\rm life}} \approx 0.586 \, ,  
\label{cconstant} 
\ee
where $N_{\rm life}$ is the number of atomic time scales required
for a functioning biosphere (and where we use $N_{\rm life}$ = 
$10^{33}$ to obtain the numerical value). 

The maximum allowed value of the stellar mass defines a maximum value
of the parameter $X$, i.e.,
\be
X \le X_{\rm max} \approx 50\,. 
\label{xmax} 
\ee
Finally, the minimum stellar mass can be written in terms of 
the minimum value of $X$ such that 
\be
X \ge X_{\rm min} = D a^{3/2} \thetacen^{-9/4} \,,
\label{xmin} 
\ee
where the constant $D$ is defined by 
\be
D = 6 \left( 3\pi \right)^{1/2} 
\left( {\pi^2 \mpro \over 5 \melect} \right)^{3/4} 
\alpha^{3/2} \approx 5.36\,. 
\label{dconstant} 
\ee

\subsection{Upper Bound on the Fine Structure Constant} 
\label{sec:alphabound} 

The bounds for a minimum stellar temperature (equation
[\ref{temperature}]) and a minimum stellar lifetime (equation
[\ref{lifetime}]) can be combined and written in the form
\be
C a^4 \thetacen > g X^2 > a^{12} \thetacen^{14} B^{-2} \,.
\ee
The outer parts of the combined inequality thus imply 
the constraint 
\be
C B^2 > a^8 \thetacen^{13} \,. 
\ee
In order for the stellar structure equation (\ref{starstruck})
to have a valid solution, the temperature parameter $\thetacen$ 
is bounded from below, i.e., 
\be
\thetacen > (\thetacen)_{\rm min} \approx 0.869 \,. 
\ee
We thus obtain the bound 
\be
a < \left( C B^2 \right)^{1/8} (\thetacen)_{\rm min}^{-13/8} 
\approx 574 \, . 
\label{abound} 
\ee

\subsection{Upper Bound on the Gravitational Constant
and Ratio of Force Strengths} 
\label{sec:gravbound} 

Next we derive an upper limit on the gravitational structure constant
$\alpha_G$ along with an upper limit on the ratio $\alpha_G/\alpha$.
If we combine the stellar structure equation (\ref{starstruck}) with
the minimum value of the stellar mass parameter from equation
(\ref{xmin}), we obtain the inequality 
\be
I(\thetacen) \thetacen^{-8} \ge A X_{\rm min} ^4 g a^{-8} =
A D^4 a^6 \thetacen^{-9} g a^{-8} \,,
\ee
which can be rewritten in the form 
\be
\thetacen I(\thetacen) \ge A D^4 g a^{-2} \,.
\ee
We also require $X_{\rm min} \le X_{\rm max}$, where this 
condition can be used to obtain a bound on $a$, i.e., 
\be
a^2 \le 50^{4/3} D^{-4/3} \thetacen^{3}  \,. 
\ee
Combining the previous two equations then yields the 
inequality 
\be
\thetacen I(\thetacen) \ge A D^4 g a^{-2} \ge 
A D^4 g 50^{-4/3} D^{4/3} \thetacen^{-3}  \,,
\ee
which can be rewritten in the form 
\be
g \le A^{-1} D^{-16/3} 50^{4/3} 
\left[ \thetacen^4 I(\thetacen) \right]_{\rm max} 
\approx 4.5 \times 10^6 
\left[ \thetacen^4 I(\thetacen) \right]_{\rm max} \,. 
\label{gbound} 
\ee
Note that we have replaced the value of the function 
$\thetacen^4 I(\thetacen)$ with is maximum value. 

Similarly, we can make an analogous argument to find a limit 
on the ratio $g/a$, which results in the upper bound
\be
{g \over a} \le A^{-1} D^{-14/3} 50^{2/3} 
\left[ \thetacen^{5/2} I(\thetacen) \right]_{\rm max} 
\approx 10^6 
\left[ \thetacen^{5/2} I(\thetacen) \right]_{\rm max} \,. 
\label{govabound} 
\ee

Using the definition (\ref{iintegral}) of the integral function
$I(\thetacen)$, we can find a bound on the function of the form
\be
I(\thetacen) < J_0 \thetacen^2 \exp[-3\thetacen] \,,
\ee
where $J_0$ is given by the integral 
\be
J_0 = \int_0^{\xi_\ast} \xi^2 d\xi f^{2n-2/3} \,,
\ee
where $f(\xi)$ is the solution to the Lane-Emden equation 
for polytropic index $n$. Note that we can also write the 
expression for $J_0$ in the form 
\be
J_0 = \int_0^{\xi_\ast} \xi^2 d\xi f^n \left[ f^{n-2/3} \right] \,.
\ee
As long as the polytropic index $n > 2/3$, the factor in square 
brackets is less than unity, whereas the remaining part of the 
expression is just $\mzero$, so that we obtain the bound 
\be
J_0 < \mzero\,.
\ee
Given the upper limit on $I(\thetacen)$, we can find an upper 
limit on functions of the form 
\be
F (\thetacen) = \thetacen^k I(\thetacen) \,,
\ee
which is bounded by 
\be
F \le F_{\rm max} < \mzero \left({k+2 \over 3}\right)^{k+2} 
\exp[-(k+2)] \,. 
\ee
Using this result to evaluate the bounds of equations 
(\ref{gbound}) and (\ref{govabound}), we find the limits 
\be
g \simless 1.9 \times 10^6
\qquad {\rm and} \qquad 
{g \over a} \simless 1.9 \times 10^5 \,. 
\ee


\end{document}